# Towards Increase in Quality by Preprocessed Source Code and Measurement Analysis of Software Applications

**Zeeshan Ahmed and Saman Majeed**
University of Wuerzburg, Germany
Authors E-mails: {zeeshan.ahmed, saman.majeed} @uni-wuerzburg.de
Web Link: www.uni-wuerzburg.de

*Abstract*— **In this paper two intensive problems faced during software application's analysis and development process arose by the software industry are briefly conversed i.e. identification of fault proneness and increase in rate of variability in the source code of traditional and product line applications. To contribute in the field of software application analysis and development, and to mitigate the aforementioned hurdles, a measurement analysis based approach is discussed in this paper. Furthermore, a prototype is developed based on the concepts of discussed approach i.e. analyzing preprocessed source code characteristics, identifying additional level of complexities using several procedural and object oriented source code measures and visualizing obtained results in different diagrams e.g. bar charts, file maps and graphs etc. Developed prototype is discussed in detail in this paper and validated by means of an experiment as well.**

## 1. Introduction

The main objective of every product based organization in any field of industry is the successful revenue generation which totally depends on the market value of manufactured product. If the product is good, qualitative and timely available in best economic price in the market, the probability of getting more profit out of it is also very high. On the other hand if the quality of the product is not up to the mark then there will be high chances that the organization/company will fall into trouble soon. Thus the maintenance of product's quality is of pivotal importance in software industry. In this article the main focus of discussion is software industry and software applications.

These days software applications have brought a revolution in the world. In almost every corporate organization and industry software (desktop and web based) applications are employed in different forms e.g. *distributed enterprise systems, database management systems, knowledge management systems, product data management systems, artificially intelligent systems, decision support systems, bioinformatics systems, satellite systems and several other special purpose systems* etc. Where software industry is contributing in almost every field of the world there at the same time it is also facing some problems due to the high failure rate.

Software applications are mainly of two kinds i.e. *Project and Product* having only one major difference in between. Software Project is an application developed meeting client's requirement e.g. a web site based on some personal information or a database system for inventory control of a particular business etc. Whereas the software Product is the special purpose system, developed by keeping some standard requirements of meeting some particular goal e.g. Operating Systems, Broadcasting TV Channel Systems and Mobile Systems etc. Likewise a major difference, there is also a commonality in between that these two different kinds of software applications follow the software development processes and life cycles.

No doubt with the passage of time software project or product development processes and life cycles have become mature enough to support the development of software applications but still there are some areas which need more attention and to be improved as well. One of the most important parts of any software development process or life cycle which can directly affect the quality of a software application is Programming i.e. about to write preprocessed source code based on particular computer understandable languages, which later (if successfully) processed by computer with the use of some particular compilers or interpreters and become in the shape of a software application. With the passage of time software applications are increasing in size and becoming large, especially when we talk about software applications based on Product Line Architectures (these are the applications based on the architecture consisting of several different applications working independently and together in an integrated form as well). This increase in size of software applications increases the (preprocessed) Line of Code (LOC), which somehow can cause an increase in the level of complexity in the source code.

If the complexity level in software application preprocessed source code cannot be properly traced and tracked during software development and debugging processes, it can start causing an increase in variabilities in the source code which can ultimately increase the probability of getting a fault prone application resulting loss in business at the end. This discussion arises two questions leading to the same goal of having a fault prone software application i.e.

1. *How to reduce the rate of variabilities in preprocessed source code of software applications, especially if it is in development process?*

2. *How to identify the rate of fault proneness if the application is already developed?*

First of all to avoid getting variabilities in source code we have to stop the increase in complexities and to stop rise in complexities in preprocessed source code we have to track the process of preprocessed source code development line by line, which can be a quite hectic and slow task for an individual. Secondly to identify the rate of fault proneness in already developed applications we have to identify the rate and the effect of variabilities in preprocessed source code.

Focusing on the aforementioned problems of reducing variabilities and identifying the rate of fault proneness in software applications we have discussed an approach in section 2 of this research paper. Following the concept of discussed approach, we present basic implementation designs, involved technologies and developed prototype in section 3 of this research paper. We validate developed prototype by analyzing a real time software application in an experiment in section 4 and conclude the discussion in section 5 of this research paper.

## 2. Approach

The solution to the aforementioned problems is composed of three integrated components i.e., *Analysis, Measurement and Visualization* respectively [6], each is responsible of performing a specific task leading towards in depth understanding of the preprocessed source code in context to the rate of variabilities and fault proneness in developed product, as shown in Figure 1.

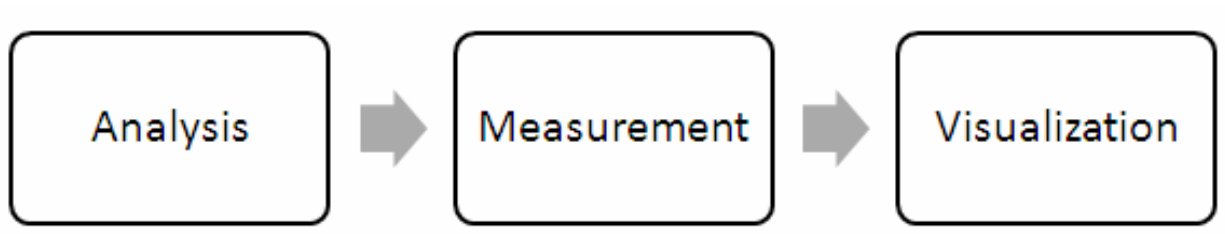

Figure.1. Three component based approach

Analysis reveals the results of the preprocessed source code to identify its complexities and variabilities. Measurement supports by calculating preprocessed source code based metrics to evaluate the rate of fault proneness. Whereas Visualization helps in understanding the complex statistical results (from Measurement) by producing different diagrams e.g. graphs, charts and maps etc.

To provide a comprehensive solution to the software practitioners in analyzing the software application's preprocessed source code to identify the level of complexities leading to variabilities by calculating metrics and identifying the rate of fault proneness, we have developed above discussed approach in the form of a software prototype application i.e. ZAJ; Java preprocessed source code analyzer. ZAJ only analysis preprocessed source of applications developed using Java Programming Language [1].

## 3. Prototype - Design and Development

ZAJ is designed to dynamically analyze the internal preprocessed source code characteristics of both the traditional and product line architecture based software applications. It computes the relevant source code metrics to identify the rate of fault proneness in software applications and visualize the obtained results during analysis and measurement. ZAJ is the enhanced version of ZAC [6], designed for the analysis of software applications developed in Java programming language whereas ZAC is designed for the analysis of the preprocessed source code of applications developed using C/C++ programming languages.

### *3.1. Work Flow*

To fulfill the desired jobs and obtain required results ZAJ is divided in to four components i.e., *Analyzer, Data Manager, Measurer, and Visualizer* which work in a cyclic order as shown in Figure 2.

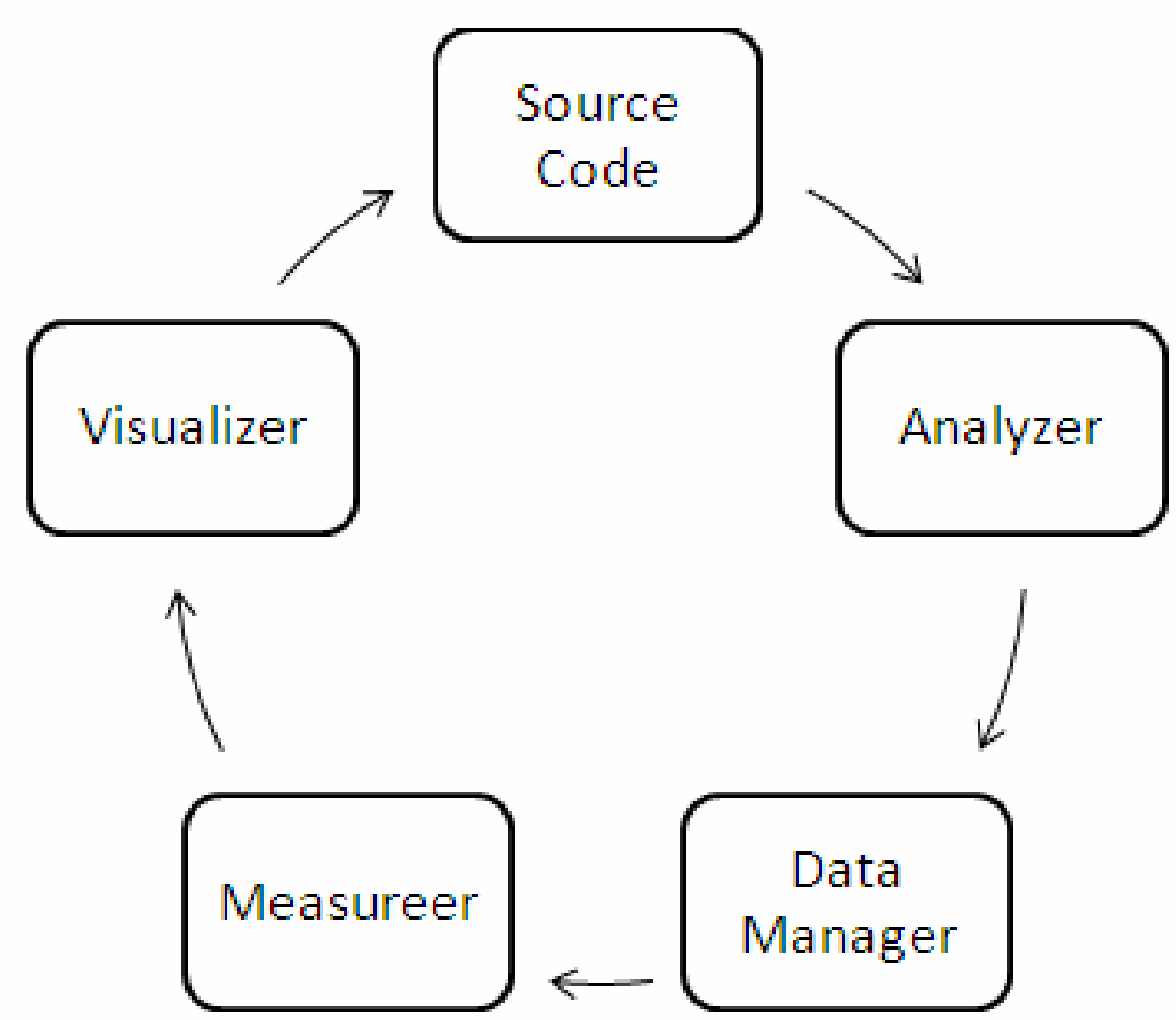

Figure. 2. Conceptual Design

ZAJ Cyclic Process works in five steps, in the first step preprocessed source code of software application developed using Java programming language is taken as input, then in the second step internal software characteristics e.g. package(s), class(es), method(s), function(s), declaration(s), expression(s) and condition(s) etc. are analyzed. Then in step 3, the resultant information is stored and managed in a relational database management system, while in step 4 using Goal Question Metrics (GQM) [2] source code metrics are calculated from the resultant output of Analyzer stored in relation database using Data Manager. In the end as the last step visualization is produced of obtained results in different diagrams e.g. graphs, line charts, bar charts and tree maps etc.

### *3.2. Internal Work Flow*

Designed internal work flow of ZAJ consists of six components .i.e., *Source Code Analyzer, Semantic Modeler, Data Manager, Measurer, Visualizer and Editor,* as shown in Figure 3. Each of six components altogether work in a sequence to attain the final goal.

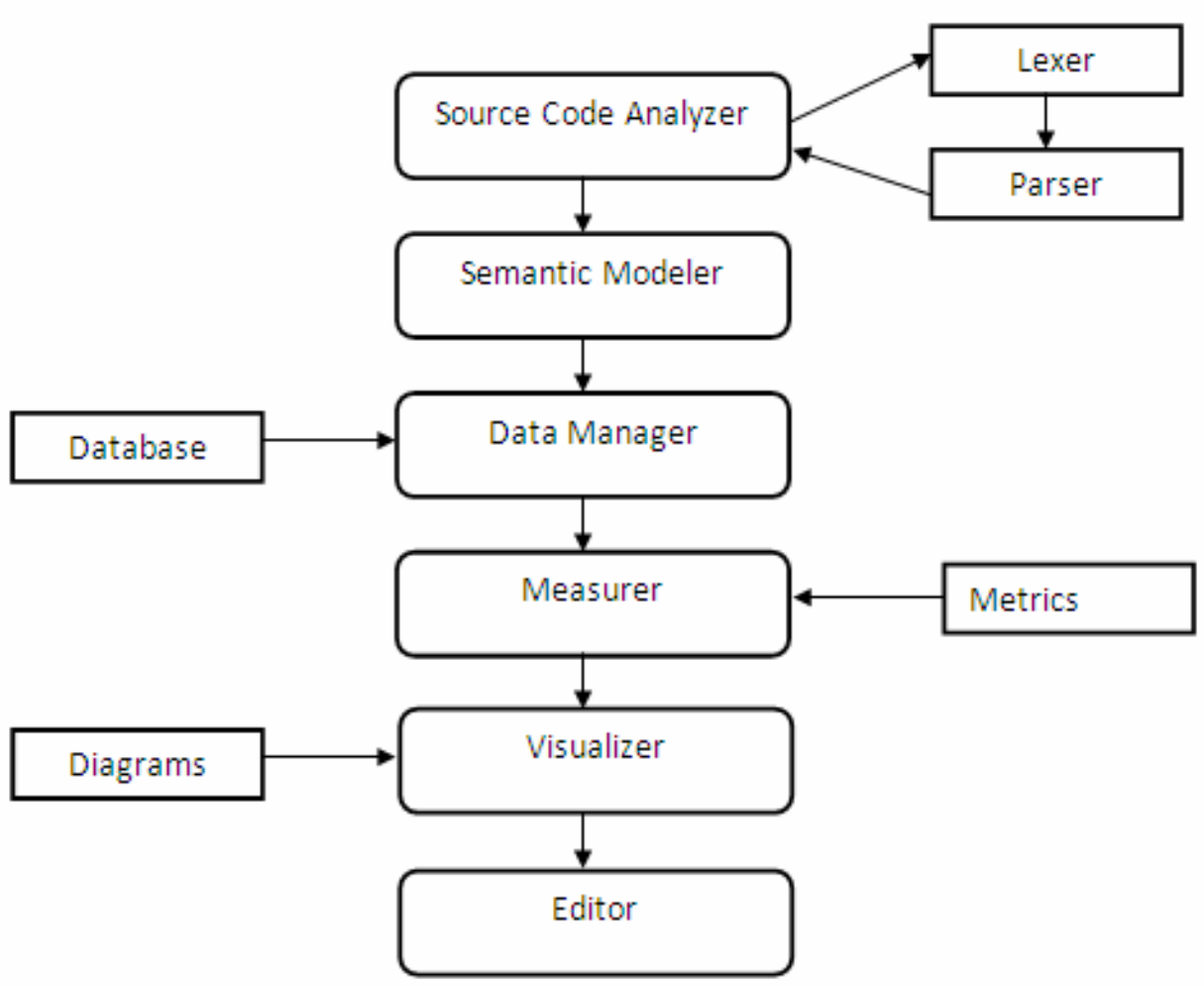

Figure. 3. Internal Work Flow

The preprocessed source code of software application is first treated by Source Code Analyzer which analyses the internal source code characteristics by dividing it into two further subcomponents i.e., *Lexer and Parser*. Lexer tokenizes the source code by generating lexical tokens and verifies the syntax with respect to the used grammar of Java programming language. Parser uses the generated tokens by lexer to verify the semantics of syntax with respect to the grammar rules of Java programming language.

Based on the resultant information, if the syntax and semantic is verified then the Semantic Modeler generates a semantic model consisting of the resultant information of Source Code Analyzer as shown in Figure 4. The designed model is based on the grammatical structure of Java programming language and consists of following three components i.e., *Application, Package, and Class*. In the first constituent, the whole preprocessed source code is classified into two further subcategories i.e., *Artifacts and Components*. Artifacts contain the information about all the project files including preprocessed source code and other files e.g. Java files, Images files etc. whereas Component contains only the preprocessed source code files.

In the second constituent Package, all the packages are included and filtered again in to two subcategories .i.e., *Libraries and Project Files*. Libraries are the third party developed preprocessed source code files e.g. APIs, used in the development of the project. Project files are newly source code files developed for the respective project implementation. Then in the last and third constituent Class again we have three more categories *Declarations Methods and Expressions*. Declarations are the variables used in preprocessed source code. Methods contain all functions and methods, whereas Expressions include mathematical statements (based on variable and value combinations) in the preprocessed source code.

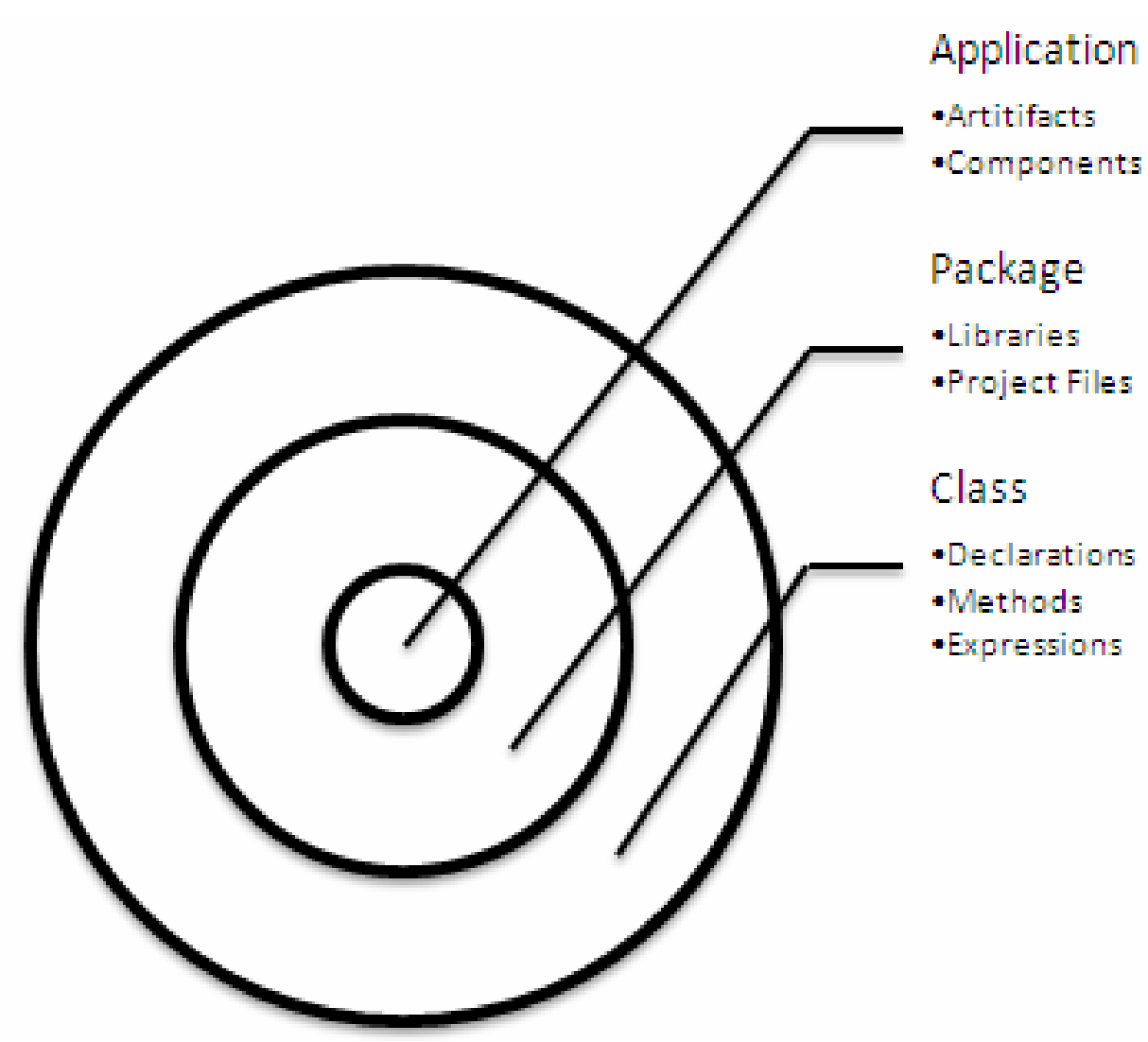

Figure. 4. Semantic Model

Constructed model is then stored and managed in relational database by Data Manger in two categories .i.e., *Relations and Relationships*. The stored relational information is then used by Measurer to calculate source code metrics which then can be visualized using Visualizer by producing different diagrams. Editor is the last component which provides options to make textual changes to extract and utilize the obtained results to take advantage in making analytical decisions by making some external statistical calculations.

### *3.3. Technologies Involved*

ZAJ is implemented using freely available tools and technologies i.e. Java (Object oriented software development language), Antlr (a language recognizer which interprets, compiles, and translates grammatical descriptions containing actions in a variety of target languages) [3], MySQL (a relational database management systems) [4] and Graphviz ( is an open source graph visualization software API used to produce visual representation of structural results in abstract graphs and networks) [5], as shown in Figure 5.

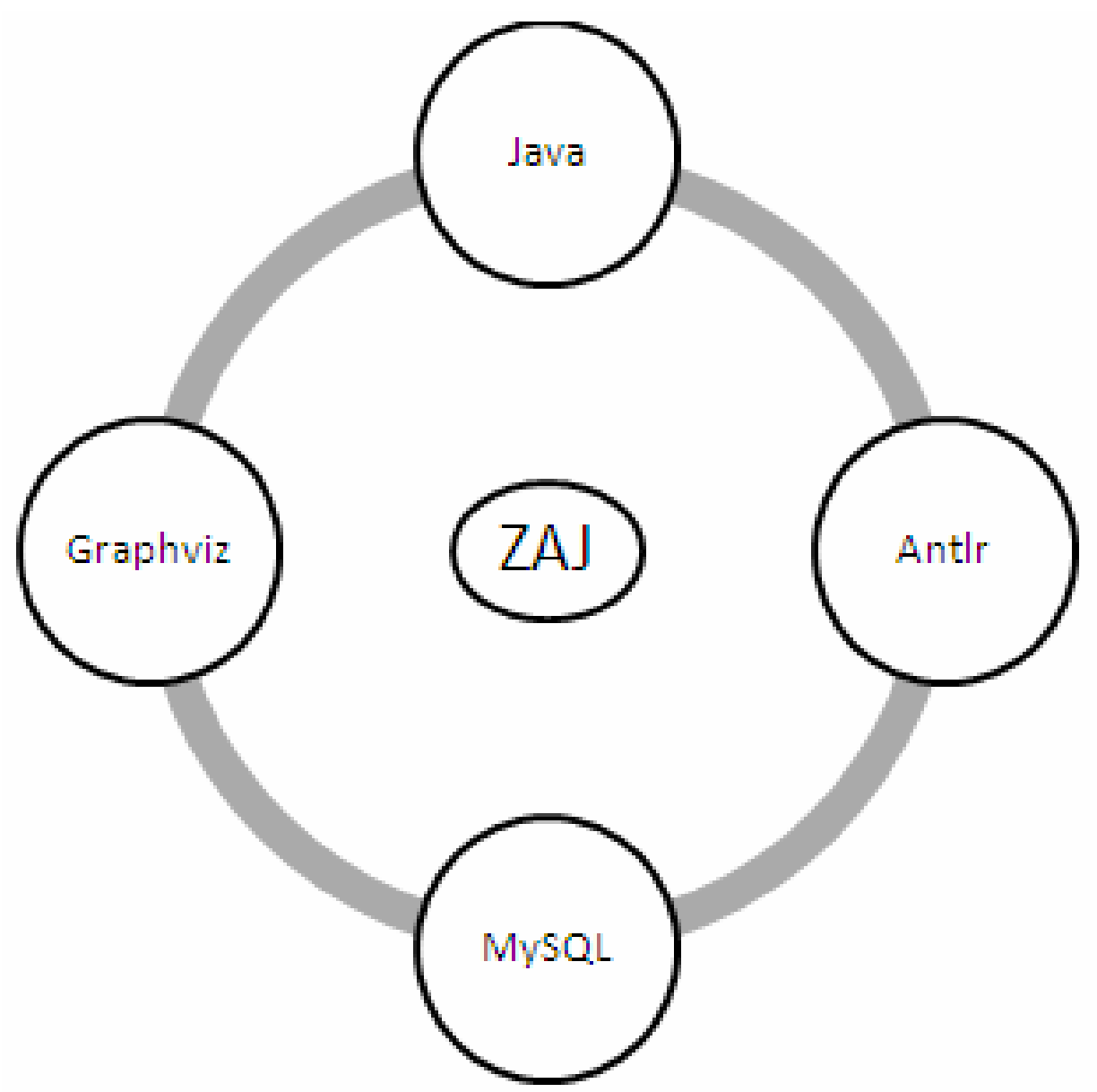

Figure. 5. Involved Technologies

### *3.4. Tool*

Using earlier discussed system designs and technologies, a small prototype with limited options is developed consisting of three modules .i.e., *Analyzer, Measurer and Visualizer*. This prototype is capable of taking preprocessed source code of software applications as input, analyzing input characteristics using Analyzer, measuring some metrics using Measurer and producing some visualizations using Visualizer. Currently developed version of prototype is capable of doing the following tasks i.e.

- Take complete traditional and product line applications along with all the relevant and irrelevant files as input.
- Identify the relevant source code files.
- Analyze preprocessed source code files by generating lexical tokens and parsing them.
- Create semantic model and store in to database.
- Calculate size, complexity and inheritance relationships described in detail in Table 1.
- Visualize the results in graphs, tree maps and charts described in Table 2

Table 1.
Calculates Size, Complexity and Inheritance relationships

| Category | Metrics |
|---|---|
| Size Measures | -Number of artifacts<br>-Number of components<br>-Number of packages<br>-Number of classes |
| Complexity Measures | -Number of control paths<br>-Number of internal level class(es) based on |
| Inheritance Measures | -Number of control paths<br>-Number of internal level class(es) based on number of methods and statements |

Table 2.
Graphs, Tree Maps and Charts

| Category | Context |
|---|---|
| Graphs | -Package relationship<br>-Class relationship |
| File Map | - Java file<br>-Class files<br>-Jar Files<br>-Artifacts |
| Charts | -Size<br>-Number of features contained in the classes |

## 4. Validation

To validate the potential and effectiveness of developed prototype we have performed a simple experiment. We have used a client server based distributed database application (based on product line architecture), developed using Java programming [1]. This application is Print Application, designed and developed for managing the number of prints taken by several users from computer labs using different printers and plotters. This application is deployed in computer labs and network administration department. The job of Print Application is to check printer, if some user sends a request for printing then it must record and maintain the information about printer, print job and user in database.

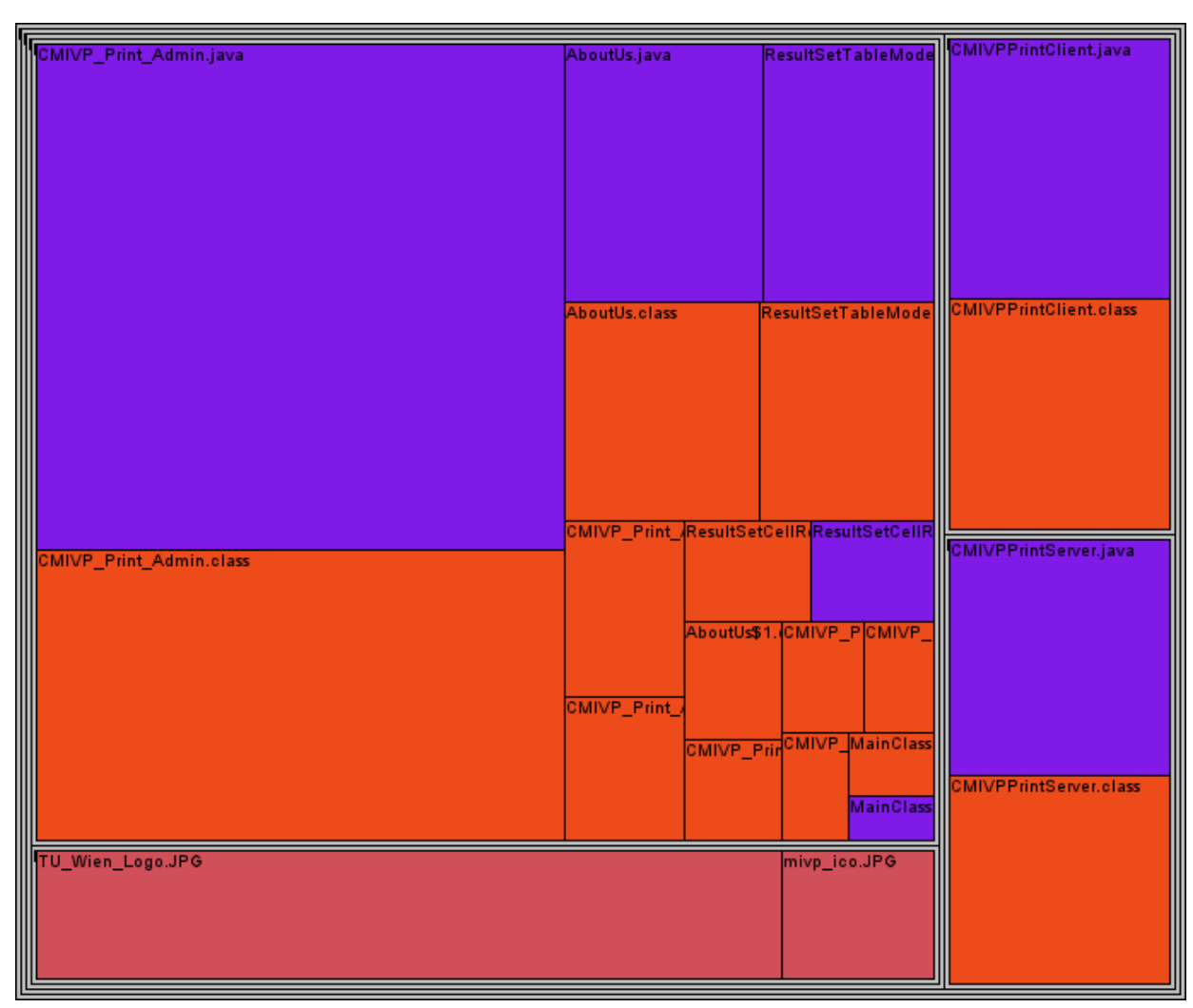

Figure.6. File Map

Moreover this application also provides inventory control system to generate bills of taken prints. We input the preprocessed source code of Print Application in developed prototype and obtained several observations; some of them are clearly described and presented in Figure 6, 7 and 8.

Figure 6 presents a file map, generated by the developed prototype after the preprocessed source code analysis of input application. This map is based on the number of files used in input application, plotted with respect to the size and classified in different colors e.g. files represented in blue boxes are Java Files, files drawn with orange color boxes are representing Class Files and files with pink color boxes are representing image Files. The size of each box is with respect to the size of the file where as the placement of each box is with respect to the placement and association of files with each other and with directory structure. This visual representation can be helpful for the software practitioners in analyzing the over all structure of the project.

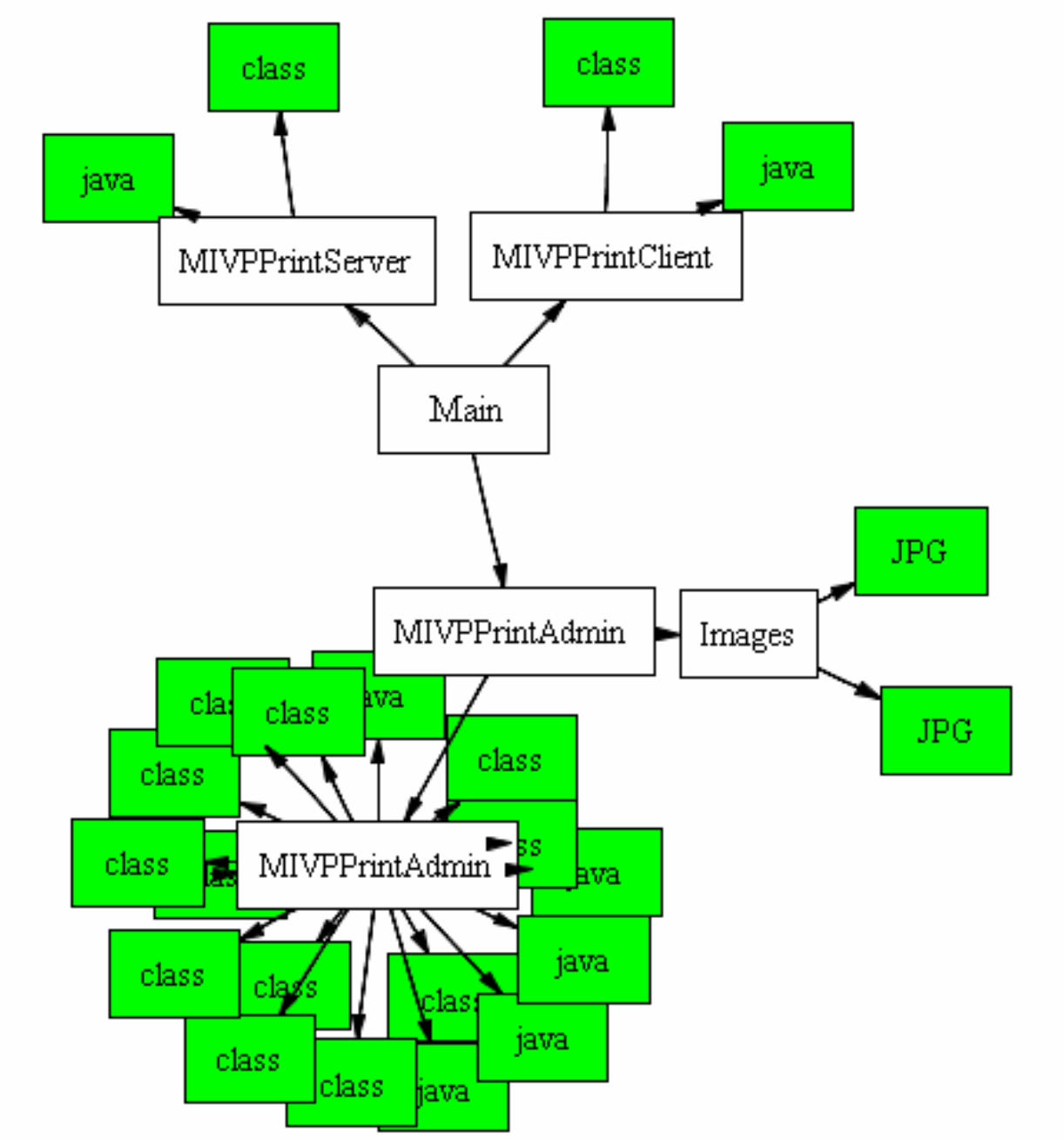

Figire.7. Package - Class relationships

As shown in Figure 7, there is a tree graph representing the overall inheritance relationships of programmed preprocessed source code based on the number of packages and classes used in the development of Print Application e.g. which package is associated or inherited with which package and which class is underdeveloped and related with which class. This kind of visual representation can be helpful for the practitioners in analyzing the overall structure of the packages and classes used in a project. Moreover software practitioners can also take advantage in analyzing the level of complexity of relationships between packages and classes as well.

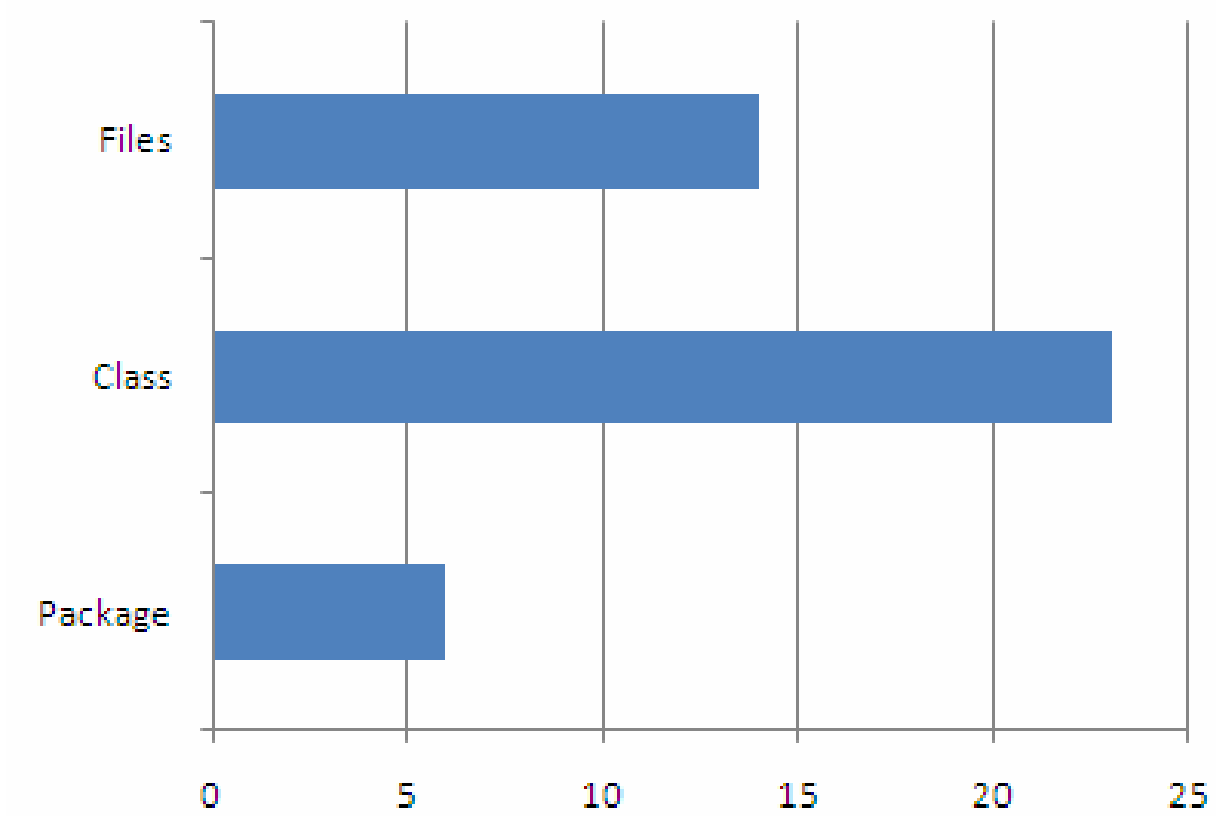

Figire.8. Bar Line Chart

As shown in Figure 8, a horizontal line chart providing the basic information about the basic source artifacts used and created in the Print Application whereas Figure 9 presents a vertical bar chart providing the information about package overview of the input project with respect to the number of class files included and size of each package. This kind of visual representation can be very helpful for the overall preprocessed source code analysis because sometimes even the rate of increase or decrease in some source code elements with respect to the class or package can also play a vital role in increasing or decreasing the complexity level.

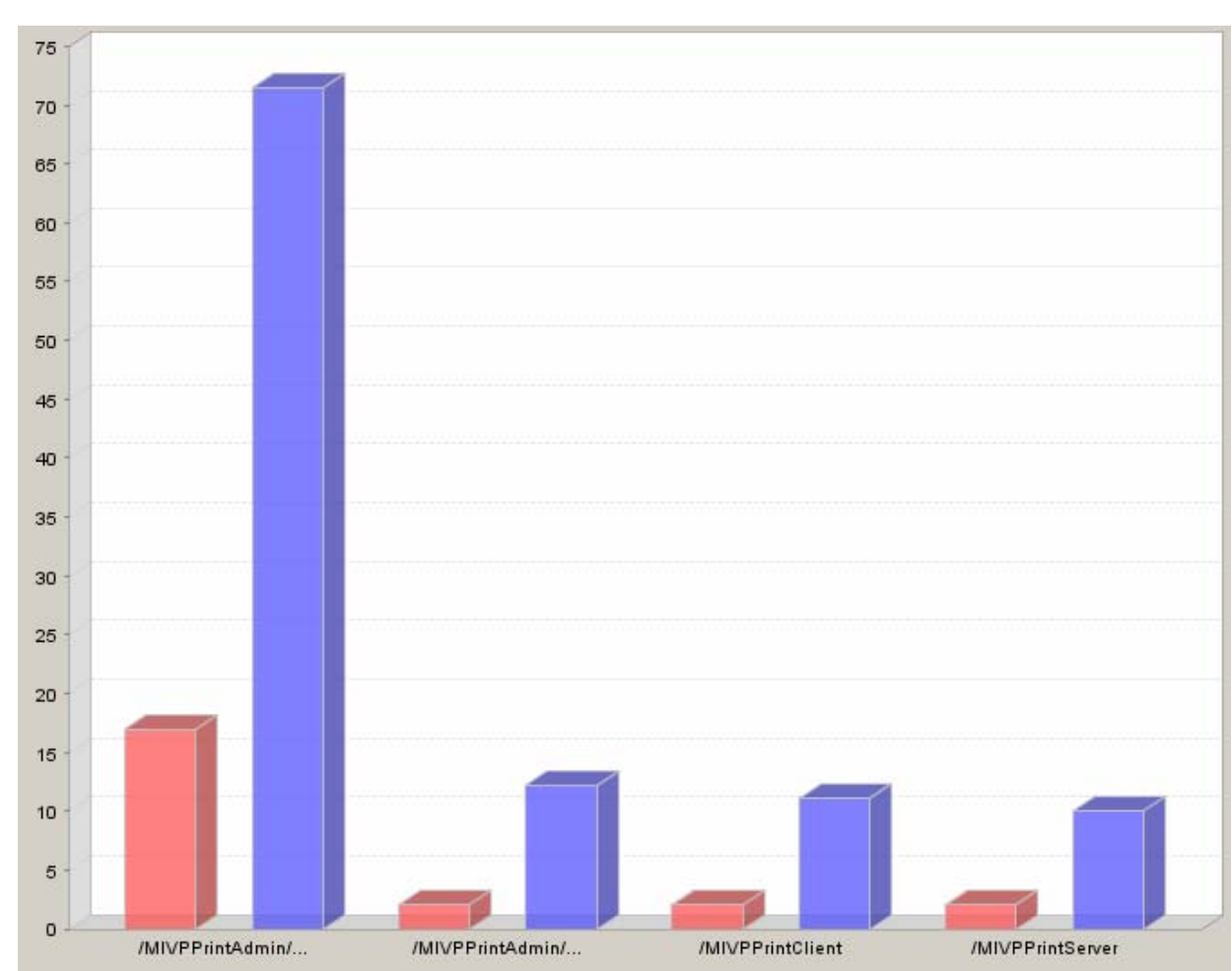

Figire.9. Bar Line Chart

## 5. Conclusion

The aim of this research was to address the issue of the identification of variabilities to reduce the rate of fault proneness in developed software applications by analyzing their preprocessed soured code. To meet aforementioned goals of this research, we have discussed the conceptual architecture of used approach in this research paper and presented a developed prototype application based on some earlier discussed implementation designs. Before concluding the discussion, we have validated the potential strength and effectiveness of developed prototype in an experiment by analyzing preprocessed source code of a product line application, the Print Application. On the basis of observed conclusions of this research work, we can say that analyzing and visualizing preprocessed source code characteristics of any traditional and product line application can be helpful in improving the quality of software applications by identifying their complex and vulnerable behaviors.

## 6. Future Recommendations

Further implementation in terms of addition of some more features in evaluating advanced software characteristics e.g. polymorphic, coupled and cohesive behavior, by analyzing preprocessed source code can be helpful for software practitioners to improve the quality of software applications.

## 7. Acknowledgement

We are thankful all to those who have helped in this research and development.

## References

[1] Sun Microsystems's production, Java, Reviewed February 2008, <www.java.com>

[2] Victor R. Basili, Gianluigi Caldieram, H. Dieter Rombach, "THE GOAL QUESTION METRIC APPROACH". *Encyclopedia of Software Engineering, John Wiley & Sons,* Inc., 1994, pp. 528-532

[3] ANTLR, Reviewed February 2008, <www.antlr.org>

[4] MySql Relational Database Management System, Reviewed February 2008, <www.mysql.com>

[5] Graphviz: Graph Visualization Software, Reviewed February 2008,

[6] Z. Ahmed,"Towards Performence Measurement and Metric based Analysis of PLA Applications", In *International Journal of Software Engineering & Applications (IJSEA)*, Vol.1, No.3, July 2010

## Author's Biography

Author; Zeeshan Ahmed, is a Research Scientist at University of Wuerzburg. He has altogether on record more than 12 years of University Education and more than 8 years of Professional Experience of working at different multinational organizations in the field of Computer Science with emphasis on software engineering of artificially intelligent applications and product line architectures.

Coauthor; Saman Majeed, is a Research Scientist at University of Wuerzburg. She has altogether on record more than 6 Years of University Education and more than 2 Years of Professional Experience of working as Researcher.